# Dissipative Quantum Systems with Potential Barrier. General Theory and Parabolic Barrier

Joachim Ankerhold[1], Hermann Grabert[1], and Gert-Ludwig Ingold[2]
[1]*Fakultät für Physik der Albert–Ludwigs–Universität, Hermann-Herder-Str. 3,
D-79104 Freiburg i. Br., Germany*
[2]*Institut für Physik, Universität Augsburg, Memmingerstr. 6,
D-86135 Augsburg, Germany*
(January 20, 1995)

## Abstract

We study the real time dynamics of a quantum system with potential barrier coupled to a heat-bath environment. Employing the path integral approach an evolution equation for the time dependent density matrix is derived. The time evolution is evaluated explicitly near the barrier top in the temperature region where quantum effects become important. It is shown that there exists a quasi-stationary state with a constant flux across the potential barrier. This state generalizes the Kramers flux solution of the classical Fokker-Planck equation to the quantum regime. In the temperature range explored the quantum flux state depends only on the parabolic approximation of the anharmonic barrier potential near the top. The parameter range within which the solution is valid is investigated in detail. In particular, by matching the flux state onto the equilibrium state on one side of the barrier we gain a condition on the minimal damping strength. For very high temperatures this condition reduces to a known result from classical rate theory. Within the specified parameter range the decay rate out of a metastable state is calculated from the flux solution. The rate is shown to coincide with the result of purely thermodynamic methods. The real time approach presented can be extended to lower temperatures and smaller damping.

# I. INTRODUCTION

Barrier penetration phenomena in quantum systems are ubiquitous in physics and chemistry [1]. Since the reaction coordinates describing the transition across the barrier typically interact with a large number of microscopic degrees of freedom, a useful theory has to start out from a formulation of quantum mechanics which incorporates the effects of a heat bath environment. In the classical region of thermally activated barrier crossings dissipation is naturally accounted for by the generalized Langevin equation or related methods. Following Kramers [2] transition rates can then be calculated from a nonequilibrium steady state solution describing a constant flux across the potential barrier.

Unfortunately, at present a correspondingly well-founded approach to dissipative barrier transmission processes is not available in the quantum regime. Clearly, the description of dissipation within the framework of quantum theory has extensively been discussed in the literature. For processes involving tunneling Caldeira and Leggett [3] have shown that a path integral formulation is particularly suitable. These studies and extensions thereof to finite temperatures by Larkin and Ovchinnikov [4] and by Grabert, Weiss, and Hänggi [5] are based on a thermodynamic method for the calculation of transition rates. In this approach pioneered by Langer [6] one determines the free energy of an unstable system by means of an analytical continuation and extracts the transition rate from the imaginary part. This technique has developed through the work of Stone [7] and Coleman and Callan [8]. A related approach based on transition state theory has been developed by Miller [9] and extended to dissipative systems by Pollak [10]. However, Hänggi and Hontscha [11] have shown explicitly that the two approaches are in fact equivalent. Analytical and numerical results on imaginary free energy calculations for dissipative metastable systems are summarized in an article by Grabert, Olschowski, and Weiss [12]. While the method was found to be very successful in explaining experimental data in well-controlled systems [13], its range of validity is not exactly known due to the lack of a first principle derivation of the rate formulas.

In this and subsequent articles we re-examine the problem of dissipative barrier penetration on the basis of a dynamical theory. The general framework is provided by a path integral description of the time evolution of the density matrix of dissipative quantum systems introduced by Feynman and Vernon [14] and extended to a wider class of useful initial conditions by Grabert, Schramm, and Ingold [15]. The theory is also presented in a recent book by Weiss [16] and a brief outline is given in section II. Thereby we also introduce the basic notation. In the remainder of this paper we apply the theory to a system with a potential barrier the height of which is large compared to other relevant energy scales. This allows for a semiclassical evaluation of the path integrals. In sections III and IV the time dependent semiclassical density matrix is determined explicitly in the high temperature region where the nonequilibrium density matrix near the barrier top is only affected by the harmonic part of the barrier potential. In section V we then show that the time evolution of the density matrix allows for a nonequilibrium flux state which is essentially stationary within a large time domain and can be shown to be a quantum generalization of the Kramers flux solution. These high temperature results are also part of the thesis by one of us [17].

In section VI we use the flux solution to derive the decay rate of a metastable state in the region of predominantly thermally activated processes. We obtain the well-known formula for quantum corrections to the classical Arrhenius rate [1,12], however, supplemented by a definite condition for the range of damping parameters within which this result is valid. Thus, the dynamic approach presented here delineates the parameter region where a purely thermodynamic rate calculation suffices. On the other hand, the flux solution cannot only be used to extract transition rates. In fact, it has already been applied to nuclear fission. There, the number of evaporated neutrons is related to the time the system needs to get from the barrier maximum to the scission point which may be derived from the flux solution [18]. In subsequent papers it will be shown that the approach presented here can be extended to lower temperatures where the potential barrier is overcome primarily by tunneling processes.

## II. PATH INTEGRAL DESCRIPTION OF DISSIPATIVE QUANTUM SYSTEMS

In this section we introduce some key quantities and formulas for quantum Brownian motion for later use. The notation follows the review by Grabert et al. [15] to which we refer for further details. We also introduce the barrier potential considered in subsequent sections and provide a formulation in terms of dimensionless quantities.

### A. Model Hamiltonian and barrier potential

We investigate the dynamics of a system with a relevant degree of freedom that may be visualized as the coordinate $q$ of a Brownian particle of mass $M$ moving in a potential field $V(q)$. The stochastic motion of the particle is due to its interaction with a heat bath. The entire system is described by the model Hamiltonian [19,3]

$$H = H_0 + H_R + H_{0R} \tag{1}$$

where

$$H_0 = \frac{p^2}{2M} + V(q) \tag{2}$$

is the Hamiltonian of the undamped particle,

$$H_R = \sum_{n=1}^{N} \frac{1}{2} \left( \frac{p_n^2}{m_n} + m_n \omega_n^2 x_n^2 \right) \tag{3}$$

describes the reservoir consisting of $N$ harmonic oscillators, and

$$H_{0R} = -q \sum_{n=1}^{N} a_n x_n + q^2 \sum_{n=1}^{N} \frac{a_n^2}{2 m_n \omega_n^2} \tag{4}$$

introduces the bilinear coupling to the coordinate $q$ of the Brownian particle. The last term in (4) compensates for the coupling induced renormalization of the potential $V(q)$. The number of the bath oscillators $N$ is assumed to be very large. To get a proper heat bath causing dissipation we will perform later the limit $N \to \infty$ with a quasi-continuous spectrum of oscillator frequencies.

Specifically, we shall consider systems where $V(q)$ has a barrier. In this paper we primarily investigate the parameter region where the barrier potential can be approximated by an inverted harmonic oscillator potential. We shall see that at lower temperatures anharmonicities of the potential field are always essential. In subsequent articles we extend the theory to lower temperatures and investigate systems with arbitrary symmetric barrier potentials. Assuming that the barrier top is at $q = 0$ and $V(0) = 0$, the barrier potential has the general form

$$V(q) = -\frac{1}{2}M\omega_0^2 q^2 \left[1 - \sum_{k=2}^{\infty} \frac{c_{2k}}{k}\left(\frac{q}{q_a}\right)^{2k-2}\right] \tag{5}$$

where the $c_{2k}$ are dimensionless coefficients and $q_a$ is a characteristic length indicating a typical distance from the barrier top at which the anharmonic part of the potential becomes essential. In particular, we assume $c_4 > 0$ so that the barrier potential becomes broader than its harmonic approximation at lower energies [1]. Also, we restrict ourselves to systems where the potential $V(q)$ does not depend on time.

A natural quantum mechanical length scale for the system near the barrier top is given by

$$q_0 = \left(\frac{\hbar}{2M\omega_0}\right)^{1/2} \tag{6}$$

which is the variance of the coordinate in the ground state of a harmonic oscillator with oscillation frequency $\omega_0$. We make the assumption that for coordinates of order $q_0$ the harmonic approximation for the potential suffices. This means that $q_a$ is large compared to the width $q_0$ and

$$\epsilon = q_0/q_a \tag{7}$$

is a small dimensionless parameter which will serve as an expansion parameter in the sequel.

### B. Path integral representation and influence functional

The time evolution of a general initial state $W_0$ of the entire system composed of the Brownian particle and the heat bath reads

$$W(t) = \exp(-iHt/\hbar) W_0 \exp(iHt/\hbar). \tag{8}$$

We shall assume that the state $W_0$ is out of thermal equilibrium due to a preparation affecting the degrees of freedom of the Brownian particle only. Then

---

[1] In fact, we may put $c_4 = 1$ thereby fixing the length scale $q_a$. To make the origin of terms in subsequent equations more transparent, we shall keep $c_4$ as a parameter, however.

$$W_0 = \sum_j O_j W_\beta O'_j, \tag{9}$$

where the operators $O_j, O'_j$ act in the Hilbert space of the particle, and

$$W_\beta = Z_\beta^{-1} \exp(-\beta H) \tag{10}$$

is the equilibrium density matrix of the entire system where the partition function $Z_\beta$ provides the normalization. In an initial state of the form (9) the system and the bath are correlated. Hence, the customary assumption that the initial density matrix $W_0$ factorizes into the density matrix of the particle and the canonical density matrix of the unperturbed heat bath is avoided. Some examples of realistic preparations leading to initial conditions of the form (9) are specified in [15]. The simplest case is an initial state which is just the equilibrium density matrix $W_\beta$. Operators $O_j, O'_j$ projecting onto a certain interval in position space provide another example.

Since we are interested in the dynamics of the particle only, the time evolution of the reduced density matrix $\rho(t) = \mathrm{tr}_R W(t)$ will be considered, where $\mathrm{tr}_R$ is the trace over the reservoir. To eliminate the environmental degrees of freedom it is convenient to employ the path integral approach [20,21]. In position representation, the equilibrium density matrix reads

$$W_\beta(\bar{q}, \bar{x}_n, \bar{q}', \bar{x}'_n) = Z_\beta^{-1} \int \mathcal{D}\bar{q}\, \mathcal{D}\bar{x}_n \exp\left(-\frac{1}{\hbar} S^{\mathrm{E}}[\bar{q}, \bar{x}_n]\right) \tag{11}$$

where the functional integral is over all paths $\bar{q}(\tau), \bar{x}_n(\tau), 0 \leq \tau \leq \hbar\beta$ with $\bar{q}(0) = \bar{q}'$, $\bar{x}_n(0) = \bar{x}'_n$, and $\bar{q}(\hbar\beta) = \bar{q}$, $\bar{x}_n(\hbar\beta) = \bar{x}_n$. The Euclidian action is given by

$$S^{\mathrm{E}}[\bar{q}, \bar{x}_n] = S_0^{\mathrm{E}}[\bar{q}] + S_{\mathrm{R}}^{\mathrm{E}}[\bar{x}_n] + S_{0\mathrm{R}}^{\mathrm{E}}[\bar{q}, \bar{x}_n] \tag{12}$$

with

$$S_0^{\mathrm{E}}[\bar{q}] = \int_0^{\hbar\beta} d\tau \left(\frac{1}{2} M \dot{\bar{q}}^2 + V(\bar{q})\right)$$

$$S_{\mathrm{R}}^{\mathrm{E}}[\bar{x}_n] = \sum_{n=1}^N \int_0^{\hbar\beta} d\tau \left(\frac{1}{2} m_n \dot{\bar{x}}_n^2 + \frac{1}{2} m_n \omega_n^2 \bar{x}_n^2\right)$$

$$S_{0\mathrm{R}}^{\mathrm{E}}[\bar{q}, \bar{x}_n] = \sum_{n=1}^N \int_0^{\hbar\beta} d\tau \left(-a_n \bar{q}\, \bar{x}_n + \bar{q}^2 \frac{a_n^2}{2 m_n \omega_n^2}\right). \tag{13}$$

The position representation of the time evolution operator $K(t) = \exp(-iHt/\hbar)$ of the entire system is

$$K(q_f, x_{n_f}, t, q_i, x_{n_i}) = \int \mathcal{D}q\, \mathcal{D}x_n \exp\left(\frac{i}{\hbar} S[q, x_n]\right). \tag{14}$$

This functional integral sums over all paths $q(s), x_n(s), 0 \leq s \leq t$ with $q(0) = q_i$, $x_n(0) = x_{n_i}$, and $q(t) = q_f$, $x_n(t) = x_{n_f}$. The action reads

$$S[q, x_n] = S_0[q] + S_{\mathrm{R}}[x_n] + S_{0\mathrm{R}}[q, x_n] \tag{15}$$

with

$$S_0[q] = \int_0^t ds \left(\frac{1}{2}M\dot{q}^2 - V(q)\right)$$

$$S_R[x_n] = \sum_{n=1}^N \int_0^t ds \left(\frac{1}{2}m_n\dot{x}_n^2 - \frac{1}{2}m_n\omega_n^2 x_n^2\right)$$

$$S_{0R}[q,x_n] = \sum_{n=1}^N \int_0^t ds \left(a_n q x_n - q^2 \frac{a_n^2}{2m_n\omega_n^2}\right). \tag{16}$$

Combining the path integral representations of the three real and imaginary time propagators in (8) and (9), one obtains for the functional integral representation of the reduced density matrix

$$\rho(q_f, q_f', t) = \frac{1}{Z} \int dq_i\, dq_i'\, d\bar{q}\, d\bar{q}'\, \lambda(q_i, \bar{q}, q_i', \bar{q}') $$
$$\times \int \mathcal{D}q\, \mathcal{D}q'\, \mathcal{D}\bar{q}\, \exp\left\{\frac{i}{\hbar}(S_0[q] - S_0[q']) - \frac{1}{\hbar}S_0^E[\bar{q}]\right\} \tilde{F}[q, q', \bar{q}] \tag{17}$$

where the functional integral is over the set of paths $q(s), q'(s), \bar{q}(\tau)$ with

$$q(0) = q_i,\ q'(0) = q_i',\ \bar{q}(0) = \bar{q}'$$

$$q(t) = q_f,\ q'(t) = q_f',\ \bar{q}(\hbar\beta) = \bar{q}$$

and where

$$\tilde{F}[q,q',\bar{q}] = \int dx_{n_f} dx_{n_i} dx_{n_i}'\, Z_R^{-1} \int \mathcal{D}x_n\, \mathcal{D}x_n'\, \mathcal{D}\bar{x}_n\, \exp\{\frac{i}{\hbar}(S_R[x_n] + S_{0R}[q,x_n]$$
$$-S_R[x_n'] - S_{0R}[q',x_n']) - \frac{1}{\hbar}(S_R^E[\bar{x}_n] + S_{0R}^E[\bar{q},\bar{x}_n])\} \tag{18}$$

is the so-called influence functional, a functional integral over all closed paths $x_n(s)$, $x_n'(s)$, $\bar{x}_n(\tau)$ of the environment with

$$x_n(t) = x_n'(t) = x_{n_f},\quad x_n(0) = \bar{x}_n(\hbar\beta) = x_{n_i},\quad \bar{x}_n(0) = x_n'(0) = x_{n_i}'.$$

$Z_R$ normalizes $\tilde{F}$ so that $\tilde{F} = 1$ for vanishing interaction, i.e. $Z_R$ is the partition function of the unperturbed bath. The new normalization factor $Z$ in (17) is given by $Z = Z_\beta/Z_R$. The position representations of the preparation operators $O_j, O_j'$ in (9) give rise to a preparation function

$$\lambda(q, \bar{q}, q', \bar{q}') = \sum_j \langle q|O_j|\bar{q}\rangle \langle \bar{q}'|O_j'|q'\rangle \tag{19}$$

describing the deviation of the initial state $W_0$ from the equilibrium state $W_\beta$.

### C. Dimensionless formulation and spectral density

Before we proceed it is convenient to introduce a dimensionless formulation. In the sequel all coordinates are scaled with respect to the quantum mechanical length scale $q_0$ introduced in (6). In particular, the scaled barrier potential then reads

$$V(q) = -\frac{1}{2}q^2 \left[1 - \sum_{k=2}^{\infty} \frac{c_{2k}}{k} \epsilon^{2k-2} q^{2k-2}\right]. \tag{20}$$

All frequencies are scaled with respect to the barrier frequency $\omega_0$ and all times with respect to $\omega_0^{-1}$. The dimensionless imaginary time interval is denoted by

$$\theta = \omega_0 \hbar \beta \tag{21}$$

and the corresponding scaled imaginary time by $\sigma$, $0 \leq \sigma \leq \theta$. Furthermore, we define sum and difference coordinates for the real time paths

$$x = q - q', \ r = (q + q')/2 \tag{22}$$

and for later purposes also

$$\bar{x} = \bar{q} - \bar{q}', \ \bar{r} = (\bar{q} + \bar{q}')/2 \tag{23}$$

for the imaginary time path.

Finally, we introduce the dimensionless spectral density of the bath oscillators

$$I(\omega) = \frac{\pi}{M\omega_0^4} \sum_{n=1}^{N} \frac{a_n^2}{2m_n\omega_n} \delta(\omega - \omega_n) \tag{24}$$

which contains all relevant information on the heat bath. With the help of the spectral density sums over the environmental oscillators may be written as integrals which is convenient when performing the limit $N \to \infty$ with a quasi-continuous spectrum of the environmental oscillators. In this limit we get a proper heat bath causing dissipation.

### D. Propagating function, effective action, and damping kernel

Now, for the harmonic oscillator model of the reservoir, which is equivalent to linear dissipation, the functional integrals occurring in the influence functional (18) can be evaluated exactly. The details of the calculation were given previously [15]. As a result one finds that the influence of the heat bath is described by an effective action containing a nonlocal damping kernel. The functional integral (17) gives for the dimensionless reduced density matrix

$$\rho(x_f, r_f, t) = \int dx_i dr_i d\bar{q}\, d\bar{q}'\ J(x_f, r_f, t, x_i, r_i, \bar{q}, \bar{q}')\, \lambda(x_i, r_i, \bar{q}, \bar{q}') \tag{25}$$

where the propagating function

$$J(x_f, r_f, t, x_i, r_i, \bar{q}, \bar{q}') = Z^{-1} \int \mathcal{D}x \, \mathcal{D}r \, \mathcal{D}\bar{q} \, \exp\left(\frac{i}{2}\Sigma[x, r, \bar{q}]\right) \tag{26}$$

depending on the action $\Sigma[x, r, \bar{q}]$ given below is a 3-fold path integral over all paths $x(s), r(s), 0 \leq s \leq t$ in real time with

$$x(0) = x_i, \; r(0) = r_i, \; x(t) = x_f, \; r(t) = r_f$$

and over all paths $\bar{q}(\sigma), 0 \leq \sigma \leq \theta$ in imaginary time with $\bar{q}(0) = \bar{q}'$, $\bar{q}(\theta) = \bar{q}$. Hence, the trajectories contributing to the propagating function are composed of two paths in real time and one in imaginary time. An entire path connects $r_f$ with $x_f$ but is interrupted since $r_i \neq \bar{q}'$ and $x_i \neq \bar{q}$, in general. These intermediate points are connected by the function $\lambda(x_i, r_i, \bar{q}, \bar{q}')$. Equation (25) determines the time evolution of the density matrix starting from the initial state

$$\rho(x_f, r_f, 0) = \int d\bar{q} \, d\bar{q}' \, \lambda(x_f, r_f, \bar{q}, \bar{q}') \rho_\theta(\bar{q}, \bar{q}'), \tag{27}$$

where $\rho_\theta = \text{tr}_R(W_\theta)$ in which $W_\theta$ is the scaled equilibrium density matrix (11) of the entire system.

The effective action in the propagating function (26) is given by [15]

$$\begin{aligned}
\Sigma[x, r, \bar{q}] = & i \int_0^\theta d\sigma \left[\frac{1}{2}\dot{\bar{q}}^2 + V(\bar{q}) + \frac{1}{2}\int_0^\theta d\sigma' k(\sigma - \sigma') \bar{q}(\sigma)\bar{q}(\sigma')\right] \\
& + \int_0^\theta d\sigma \int_0^t ds \, K^*(s - i\sigma) \, \bar{q}(\sigma)x(s) \\
& + \int_0^t ds \, [\dot{x}\dot{r} - V(r + x/2) + V(r - x/2) - r_i\gamma(s)x(s)] \\
& - \int_0^t ds \left[\int_0^s ds' \gamma(s - s') x(s)\dot{r}(s') - \frac{i}{2}\int_0^t ds' K'(s - s') x(s)x(s')\right].
\end{aligned} \tag{28}$$

Here, $*$ denotes complex conjugation. The kernel $K(s - i\sigma)$ may be written as

$$K(s - i\sigma) = K'(s - i\sigma) + iK''(s - i\sigma). \tag{29}$$

The real part of $K(s - i\sigma)$ is given by

$$K'(s - i\sigma) = \frac{1}{\theta} \sum_{n=-\infty}^{\infty} g_n(s) \exp(i\nu_n \sigma) \tag{30}$$

with the Fourier coefficients

$$g_n(s) = 2 \int_0^\infty \frac{d\omega}{\pi} I(\omega) \frac{\omega}{\omega^2 + \nu_n^2} \cos(\omega s) \tag{31}$$

where the

$$\nu_n = \frac{2n\pi}{\theta} \tag{32}$$

are dimensionless Matsubara frequencies scaled with $\omega_0$. The imaginary part of $K(s - i\sigma)$ reads

$$K''(s - i\sigma) = \frac{1}{\theta} \sum_{n=-\infty}^{\infty} i f_n(s) \exp(i\nu_n \sigma) \qquad (33)$$

with the Fourier coefficients

$$f_n(s) = 2 \int_0^\infty \frac{d\omega}{\pi} I(\omega) \frac{\nu_n}{\omega^2 + \nu_n^2} \sin(\omega s). \qquad (34)$$

Furthermore, the so-called damping kernel

$$\gamma(s) = 2 \int_0^\infty \frac{d\omega}{\pi} \frac{I(\omega)}{\omega} \cos(\omega s) \qquad (35)$$

determines the imaginary part of the kernel (29) for purely real times

$$K''(s) = \frac{1}{2} \frac{d\gamma(s)}{ds}. \qquad (36)$$

For purely imaginary times $0 \leq \sigma \leq \theta$ the kernel in the Euclidian part of the effective action (28) is given by

$$k(\sigma) = -K(-i\sigma) + \gamma(0) : \delta(\sigma) :$$
$$= \frac{2}{\theta} \sum_{n=1}^{\infty} \zeta_n \cos(\nu_n \sigma) \qquad (37)$$

where we have introduced the periodically repeated delta function

$$: \delta(\sigma) := \sum_{n=-\infty}^{\infty} \delta(\sigma - n\theta). \qquad (38)$$

The coefficients

$$\zeta_n = 2 \int_0^\infty \frac{d\omega}{\pi} \frac{I(\omega)}{\omega} \frac{\nu_n}{\omega^2 + \nu_n^2}$$
$$= |\nu_n| \hat{\gamma}(|\nu_n|) \qquad (39)$$

are related to the Laplace transform $\hat{\gamma}(z)$ of the damping kernel $\gamma(s)$. The kernel $k(\sigma)$ satisfies

$$\int_0^\theta d\sigma\, k(\sigma) = 0. \qquad (40)$$

We note that all quantities describing the effect of the heat bath can be expressed in terms of the spectral density $I(\omega)$ defined in (24).

## III. SEMICLASSICAL APPROXIMATION AND PARABOLIC BARRIER

For an anharmonic potential field the 3-fold path integral (26) cannot be evaluated exactly. In the following we study the semiclassical approximation which will be seen to suffice for small $\epsilon$ and coordinates near the barrier top. For high temperatures the anharmonic terms in the potential (20) give only small corrections and a simple semiclassical approximation for a parabolic barrier with Gaussian fluctuations is appropriate. At the beginning of this section we specify the equations of motion for the minimal action paths of the effective action. In subsection III B the extremal imaginary time path for high temperatures is determined, and in subsection III C the extremal real time paths are evaluated using $\epsilon$ as an expansion parameter. The minimal effective action and the fluctuations around the extremal paths are then determined in section IV leading to the semiclassical time dependent density matrix for high temperatures.

### A. Minimal action paths

Variation of the effective action $\Sigma[x, r, \bar{q}]$ introduced in (28) with respect to $\bar{q}$ leads to the equation of motion for the minimal action path in imaginary times

$$\ddot{\bar{q}} - \int_0^\theta \mathrm{d}\sigma'\, k(\sigma - \sigma')\bar{q}(\sigma') - \frac{\mathrm{d}V(\bar{q})}{\mathrm{d}\bar{q}} = -i \int_0^t \mathrm{d}s\, K^*(s - i\sigma) x(s). \tag{41}$$

The inhomogeneity on the right hand side couples $\bar{q}(\sigma)$ to the real time motion. Variation of the effective action $\Sigma[x, r, \bar{q}]$ with respect to $x$ and $r$ leads to the equations of motion for the minimal action paths in real time

$$\ddot{r} + \frac{\mathrm{d}}{\mathrm{d}s}\int_0^s \mathrm{d}s'\, \gamma(s - s') r(s') + \frac{1}{2}\frac{\mathrm{d}}{\mathrm{d}r}\left\{V(r + x/2) + V(r - x/2)\right\} = \\ i \int_0^t \mathrm{d}s'\, K'(s - s') x(s') + \int_0^\theta \mathrm{d}\sigma\, K^*(s - i\sigma) \bar{q}(\sigma). \tag{42}$$

and

$$\ddot{x} - \frac{\mathrm{d}}{\mathrm{d}s}\int_s^t \mathrm{d}s'\, \gamma(s' - s) x(s') + 2\frac{\mathrm{d}}{\mathrm{d}x}\left\{V(r + x/2) + V(r - x/2)\right\} = 0. \tag{43}$$

The above formulae (25)–(28) for the density matrix and the equations of motion (41)–(43) for the minimal action paths hold for any potential field [15]. In the following we shall consider explicitly the barrier potential introduced in (5) and (20).

### B. Extremal imaginary time path for high temperatures

Since we seek a solution of (41) in the finite time interval $0 \leq \sigma \leq \theta$, it is convenient to employ the Fourier series expansion

$$\bar{q}(\sigma) = \frac{1}{\theta} \sum_{n=-\infty}^{\infty} q_n \exp(i\nu_n \sigma). \tag{44}$$

This series for $\bar{q}(\sigma)$ continues the path outside the interval $0 \leq \sigma \leq \theta$ as a periodic path with period $\theta$. The continuation causes jump singularities in $\bar{q}(\sigma)$ and $\dot{\bar{q}}(\sigma)$ at the endpoints which must be taken into account when calculating time derivatives of $\bar{q}(\sigma)$. In the following we shall show selfconsistently that for endpoints $\bar{q}, \bar{q}'$ of the imaginary time path $\bar{q}(\sigma)$ and endpoints $x_i$, $x_f$ of the real time path $x(s)$ that are of order 1 or smaller, the $q_n$ are of order 1 or smaller for high temperatures. Then, the potential (20) may be approximated by

$$V(\bar{q}) = -\frac{1}{2}\bar{q}^2 + \mathcal{O}(\epsilon^2). \tag{45}$$

Inserting (44) into (41) yields the Fourier representation of the equation of motion

$$(\nu_n^2 + \zeta_n - 1)q_n = i\nu_n \bar{x} - b + if_n[x(s)] + ig_n[x(s)]. \tag{46}$$

The first term on the right hand side, $i\nu_n \bar{x}$, corresponds to a $:\delta(\sigma):$ singularity of $\bar{q}(\sigma)$ according to $\bar{q}(0^-) - \bar{q}(0^+) = \bar{x}$. The coefficient $b$ is related to a $:\delta(\sigma):$ singularity of $\dot{\bar{q}}(\sigma)$

$$b = \dot{\bar{q}}(0^+) - \dot{\bar{q}}(0^-) = \dot{\bar{q}}(0^+) - \dot{\bar{q}}(\theta^-). \tag{47}$$

The functionals

$$f_n[x(s)] = \int_0^t ds\, f_n(s)x(s) \tag{48}$$

and

$$g_n[x(s)] = \int_0^t ds\, g_n(s)x(s) \tag{49}$$

describe the coupling to the real time motion. The solution of (46) reads

$$q_n = u_n\, (i\nu_n \bar{x} - b + if_n[x(s)] + ig_n[x(s)]) \tag{50}$$

with the abbreviation

$$u_n = 1/(\nu_n^2 + \zeta_n - 1) = 1/(\nu_n^2 + |\nu_n|\hat{\gamma}(|\nu_n|) - 1). \tag{51}$$

The coefficient $b$ must be determined such that the path $\bar{q}(\sigma)$ satisfies the boundary conditions $\bar{q}(0^+) = \bar{q}'$, $\bar{q}(\theta^-) = \bar{q}$. Due to the discontinuities of the periodically continued path at the endpoints, care must be taken in performing the limit $\sigma \to 0$. From (50) and (51) we obtain

$$\bar{q}(\sigma) = -\frac{1}{\theta} \sum_{n=-\infty}^{\infty}{}' \frac{\bar{x}}{i\nu_n} \exp(i\nu_n \sigma) + \frac{1}{\theta} \sum_{n=-\infty}^{\infty}{}' \frac{\bar{x}}{i\nu_n} u_n(\zeta_n - 1) \exp(i\nu_n \sigma)$$

$$- \frac{1}{\theta} \sum_{n=-\infty}^{\infty} u_n(b - if_n[x(s)] - ig_n[x(s)]) \exp(i\nu_n \sigma) \tag{52}$$

where the prime denotes the sum over all elements except $n = 0$. The first sum gives in the limit $\sigma \to 0^{\pm}$

$$\lim_{\sigma \to 0^{\pm}} \frac{1}{\theta} \sum_{n=-\infty}^{\infty}{}' \frac{\bar{x}}{i\nu_n} \exp(i\nu_n \sigma) = \pm \frac{\bar{x}}{2}. \tag{53}$$

The second sum in (52) is regular in the limit $\sigma \to 0$ and vanishes. The third sum is also regular and one obtains

$$\lim_{\sigma \to 0^{\pm}} \bar{q}(\sigma) = -b\Lambda + \frac{i}{\theta} \sum_{n=-\infty}^{\infty} u_n \, g_n[x(s)] \mp \frac{\bar{x}}{2} \tag{54}$$

where

$$\Lambda = \frac{1}{\theta} \sum_{n=-\infty}^{\infty} u_n = -\frac{1}{\theta} + \frac{2}{\theta} \sum_{n=1}^{\infty} u_n, \tag{55}$$

and where the relations $g_{-n}(s) = g_n(s)$ and $f_{-n}(s) = -f_n(s)$ have been used with $g_n(s)$, $f_n(s)$ given in (31) and (34). We note that for a harmonic oscillator $\Lambda$ is related to the variance of position [15]. For a parabolic barrier, however, $\Lambda$ has no obvious physical meaning since $\Lambda < 0$ for high temperatures. Taking into account the boundary conditions for the periodically continued path $\bar{q}(\sigma)$, we obtain from (54)

$$b = -\frac{1}{\Lambda}\left(\bar{r} - \frac{i}{\theta} \sum_{n=-\infty}^{\infty} u_n \, g_n[x(s)]\right) \tag{56}$$

with $\bar{r}$ defined in (23). Finally, using (50) and (56), the Fourier coefficients $q_n$ of the imaginary time trajectory read

$$q_n = u_n \left[i\nu_n \bar{x} + if_n[x(s)] + ig_n[x(s)] + \frac{\bar{r}}{\Lambda} - \frac{i}{\theta \Lambda} \sum_{m=-\infty}^{\infty} u_m \, g_m[x(s)]\right] \tag{57}$$

which determine the minimal action path $\bar{q}(\sigma)$ as a function of the endpoints for high enough temperatures. This result depends on the as yet undetermined real time path $x(s)$. Below we shall show that $x(s)$ remains of order 1 for all $0 \leq s \leq t$. Thus, (57) confirms our assumption made at the beginning of this subsection that for endcoordinates at most of order 1 the path $\bar{q}(\sigma)$ always remains in the vicinity of the top and is also at most of order 1.

When the temperature is lowered, i.e. the dimensionless inverse temperature $\theta$ is increased, $|\Lambda|$ becomes smaller and vanishes for the first time at a critical inverse temperature $\theta_c$

$$\Lambda(\theta_c) = -\frac{1}{\theta} + \frac{2}{\theta} \sum_{n=1}^{\infty} u_n \bigg|_{\theta=\theta_c} = 0. \tag{58}$$

For vanishing damping one has $\Lambda = -\frac{1}{2}\cot(\theta/2)$ and therefore $\theta_c = \pi$, i.e. $T_c = \hbar\omega_0/\pi k_{\mathrm{B}}$ in dimensional units. As a consequence of (58) the Fourier coefficients $q_n$ in (57) diverge for a purely harmonic barrier when $\theta_c$ is approached. This divergence corresponds to the problem of caustics for a harmonic oscillator [21,22]. Since the harmonic potential is always an approximation, anharmonic terms of the barrier potential must be taken into account for temperatures near $T_c$ even for coordinates near the barrier top. Hence, the analysis presented so far is limited to temperatures above the critical temperature $T_c$. In an earlier work [22] we have investigated the equilibrium density matrix near the top of a weakly anharmonic potential barrier for vanishing damping down to temperatures slightly below $T_c$. In the dynamical case with damping the imaginary time part of the propagating function can be calculated in a similiar way [23].

### C. Real time paths

Let us first consider the equation of motion (42) for the real time path $r(s)$. The inhomogeneity reads

$$i \int_0^t \mathrm{d}s' K'(s-s')x(s') + \int_0^\theta \mathrm{d}\sigma K^*(s-i\sigma)\bar{q}(\sigma) = i \int_0^t \mathrm{d}s' R(s,s')x(s') + F(s). \tag{59}$$

Here, we have inserted the result (52) for $\bar{q}(\sigma)$ and made use of (57) to obtain the right hand side where

$$R(s,s') = K'(s-s') + \frac{1}{\theta}\sum_{n=-\infty}^{\infty} u_n \left[g_n(s)g_n(s') - f_n(s)f_n(s')\right] \tag{60}$$

and

$$F(s) = \frac{1}{\Lambda}\bar{r}C_1(s) - i\bar{x}C_2(s) - \frac{i}{\Lambda}C_1(s)\int_0^t \mathrm{d}s' C_1(s')x(s') \tag{61}$$

with

$$C_1(s) = \frac{1}{\theta}\sum_{n=-\infty}^{\infty} u_n g_n(s)$$

$$C_2(s) = \frac{1}{\theta}\sum_{n=-\infty}^{\infty} \nu_n u_n f_n(s). \tag{62}$$

In the following we will show that for coordinates $\bar{x}, \bar{r}$ of the imaginary time path $\bar{q}(\sigma)$ and endpoints $x_i, x_f$ of the real time path $x(s)$ that are at most of order 1, the real time path $r(s)$ remains for high temperatures also inside this region provided the endpoints $r_i, r_f$ are also at most of order 1. Anharmonic terms in (42) can then be neglected and we have

$$\frac{1}{2}\frac{\mathrm{d}}{\mathrm{d}r}\left\{V(r+x/2) + V(r-x/2)\right\} = -r + \mathcal{O}(\epsilon^2). \tag{63}$$

Hence, the solution for $r(s)$ is straightforward. It has been shown elsewhere [15] that for a harmonic potential it suffices to solve the equation of motion for $r(s)$ only for the real part. With $r(s) = r'(s) + ir''(s)$ we have from (42)

$$\ddot{r}' + \frac{\mathrm{d}}{\mathrm{d}s}\int_0^s \mathrm{d}s'\gamma(s-s')r'(s') - r' = F' \tag{64}$$

where $F'$ denotes the real part of $F$ given in (61). We introduce the propagator $G_+(s)$ of the homogeneous equation with the initial conditions $G_+(0)=0$, $\dot{G}_+(0)=1$ which has the Laplace transform

$$\hat{G}_+(z) = \left(z^2 + z\hat{\gamma}(z) - 1\right)^{-1}. \tag{65}$$

The solution of (64) is then obtained as

$$r'(s) = r_f\frac{G_+(s)}{G_+(t)} + r_i\left[\dot{G}_+(s) - \frac{G_+(s)}{G_+(t)}\dot{G}_+(t)\right]$$
$$- \int_0^s \mathrm{d}s'\, G_+(s-s')F'(s') - \frac{G_+(s)}{G_+(t)}\int_0^t \mathrm{d}s'\, G_+(t-s')F'(s'). \tag{66}$$

Hence, for endpoints of order 1 or smaller the real time path $r(s)$ is also of order 1 or smaller for $0 \leq s \leq t$ and high temperatures as assumed above. When the inverse temperature is increased $F'(s) = \bar{r}C_1(s)/\Lambda$ grows and diverges at $\theta = \theta_c$ due to the vanishing of $\Lambda$. Therefore, the solution (66) becomes invalid and anharmonicities in the equation of motion (42) commence to be important for temperatures near $T_c$.

The equation of motion for $x(s)$ is homogeneous and can be shown to be the backward equation of the equation of motion for $r(s)$ for vanishing inhomogeneity [15]. Correspondingly, the solution of (43) reads

$$x(s) = x_i\frac{G_+(t-s)}{G_+(t)} + x_f\left(\dot{G}_+(t-s) - \frac{G_+(t-s)}{G_+(t)}\dot{G}_+(t)\right). \tag{67}$$

This confirms that for endpoints $x_i$, $x_f$ of order 1 or smaller $x(s)$ is at most of order 1 as assumed above. Nonlinear terms in the equation of motion (43) are at most of order $\epsilon^2$ and can be neglected.

## IV. SEMICLASSICAL DENSITY MATRIX

Having evaluated the minimal action paths, we may determine the density matrix in the semiclassical approximation by expanding the functional integral about the minimal action paths. In subsection IV A we first calculate the minimal effective action and in subsection IV B we determine the contribution of the fluctuations about the minimal action paths. It will be seen that above $T_c$ a simple Gaussian approximation for the path integral over the fluctuations about the real time and imaginary time paths suffices.

### A. Minimal effective action

Inserting the minimal action path in imaginary time (52) and (57) as well as the minimal action paths in real time (66) and (67) into the effective action (28), one gains after a straightforward but tedious calculation for the minimal effective action for inverse temperatures below $\theta_c$ the result

$$\Sigma(x_f, r_f, t, x_i, r_i, \bar{x}, \bar{r}) = \Sigma_\theta(\bar{x}, \bar{r}) + \Sigma_t(x_f, r_f, t, x_i, r_i, \bar{x}, \bar{r}). \tag{68}$$

Here,

$$\Sigma_\theta(\bar{x}, \bar{r}) = i\frac{\bar{r}^2}{2\Lambda} + i\frac{\Omega}{2}\bar{x}^2 \tag{69}$$

is the well-known minimal imaginary-time action of a damped inverted harmonic oscillator where

$$\Omega = \frac{1}{\theta} \sum_{n=-\infty}^{\infty} u_n \left(|\nu_n|\hat{\gamma}(|\nu_n|) - 1\right). \tag{70}$$

We note that for a harmonic oscillator $\Omega$ corresponds to the variance of the momentum while for a barrier there is no obvious physical meaning as it is the case with $\Lambda$.

Since the time dependence of the minimal action paths in real time is determined essentially by the dynamics at a parabolic barrier, it is advantageous to introduce the functions

$$A(t) = -\frac{1}{2}\Theta(t)\, G_+(t) \tag{71}$$

where $\Theta(t)$ denotes the step function and

$$S(t) = \Lambda \dot{G}_+(t) + G_+(t)\, C_1^+(t) \tag{72}$$

with

$$C_1^+(t) = \int_0^t ds\, C_1(s) \frac{G_+(t-s)}{G_+(t)} \tag{73}$$

through which $\Sigma_t(x_f, r_f, t, x_i, r_i, \bar{x}, \bar{r})$ can be expressed conveniently. We note, that for a harmonic oscillator $A(t)$ is the imaginary part and $S(t)$ the real part of the time dependent position autocorrelation function [15]. With (71) and (72) one has

$$\begin{aligned}
\Sigma_t&(x_f, r_f, t, x_i, r_i, \bar{x}, \bar{r}) = \\
&(x_f r_f + x_i r_i)\frac{\dot{A}(t)}{A(t)} + x_i r_f \frac{1}{2A(t)} - 2x_f r_i \left(\ddot{A}(t) - \frac{\dot{A}(t)^2}{A(t)}\right) \\
&+\bar{r}\, x_i \left(-\frac{\dot{A}(t)}{A(t)} - \frac{S}{2\Lambda A(t)}\right) + \bar{r}\, x_f \left[2\left(\ddot{A}(t) - \frac{\dot{A}(t)^2}{A(t)}\right) + \frac{\dot{S}}{\Lambda} - \frac{S\, \dot{A}(t)}{\Lambda\, A(t)}\right] \\
&+i\bar{x}x_i \left(-\Omega + \frac{\dot{S}}{2A(t)}\right) - i\bar{x}x_f \left(\ddot{S}(t) - \frac{\dot{A}(t)}{A(t)}\dot{S}(t)\right) \\
&+\frac{i}{2}x_i^2 \left[\Omega - \frac{\dot{S}}{A(t)} + \frac{\Lambda}{4A(t)^2}\left(1 - \frac{S(t)^2}{\Lambda^2}\right)\right] \\
&+ix_i x_f \left[\ddot{S}(t) - \frac{\dot{A}(t)}{A(t)}\dot{S}(t) - \frac{\Lambda}{2A(t)^2}\left\{\dot{A}(t)\left(\frac{S(t)^2}{\Lambda^2} - 1\right) - A(t)\frac{S(t)\dot{S}(t)}{\Lambda^2}\right\}\right] \\
&+\frac{i}{2}x_f^2 \left[\Omega + \Lambda\frac{\dot{A}(t)^2}{A(t)^2} - \frac{1}{\Lambda}\left(\dot{S}(t) - \frac{\dot{A}(t)}{A(t)}S(t)\right)^2\right]. \tag{74}
\end{aligned}$$

## B. Quantum Fluctuations

With the minimal actions (69) and (74) we have found the leading order term of the path integral for the propagating function (26). The path integral now reduces to integrals over periodic paths $\xi(0) = \xi(t) = 0$ and $\xi'(0) = \xi'(t) = 0$ in real time and $y(0) = y(\theta) = 0$ in imaginary time describing the quantum fluctuations around the minimal action paths.

Firstly, let us consider the real time fluctuations. The contribution of the quantum fluctuations around the extremal real time paths can be evaluated in the simple semiclassical approximation taking into account Gaussian fluctuations. In the corresponding second order variational operator the anharmonicities of the barrier potential are at most of order $\epsilon^2$ and can therefore be neglected. The linear coupling term with the imaginary time path leads only to a shift of the real time fluctuations. The second order variational operator is then simply given by that of an inverted harmonic oscillator. Using the propagator $A(t)$ introduced in (65), the contribution of the Gaussian fluctuations around each of the two real time paths $q(s)$ and $q'(s)$ reads [15]

$$\int \mathcal{D}\xi \ \exp\left[\frac{i}{4}\int_0^t \mathrm{d}s\, \xi(s)\left(\ddot{\xi}(s) + \int_0^s \mathrm{d}u\, \gamma(s-u)\,\dot{\xi}(u) - \xi(s)\right)\right] = \frac{1}{\sqrt{8\pi|A(t)|}}. \tag{75}$$

For high temperatures for which the harmonic approximation for the potential suffices, the contribution of the quantum fluctuations to the imaginary time path integral can also be approximated by the simple semiclassical approximation. Anharmonicities in the second order variational operator can then be neglected. Expanding the quantum fluctuations into a Fourier series

$$y(\sigma) = \frac{1}{\theta}\sum_{n=-\infty}^{\infty} y_n \exp(i\nu_n \sigma) \tag{76}$$

and taking into account the boundary conditions $y(0) = y(\theta) = 0$, one obtains the well-known result [15]

$$\int \mathcal{D}y \ \exp\left[-\frac{1}{4}\int_0^\theta \mathrm{d}\sigma\, y(\sigma)\left(-\ddot{y}(\sigma) + \int_0^\theta \mathrm{d}\sigma'\, k(\sigma-\sigma')\,y(\sigma') - y(\sigma)\right)\right] =$$
$$\frac{1}{\sqrt{-\Lambda}}\frac{1}{\sqrt{4\pi\theta^2}}\prod_{n=1}^{\infty} \nu_n^2\, u_n. \tag{77}$$

When the inverse temperature is increased, the simple semiclassical approximation (77) for the imaginary time path integral diverges for inverse temperatures near $\theta_c$ ($\Lambda \to 0$). This corresponds to the problem of caustics in a harmonic oscillator potential as already discussed in subsection III B. In the harmonic approximation the amplitude of the fluctuation mode with eigenvalue $\Lambda$ of the second order variational operator becomes arbitrarily large for temperatures near $T_c$. As a consequence, the Gaussian approximation breaks down near $T_c$ and anharmonic terms of the barrier potential must be taken into account for the temperature region near and below $T_c$. For vanishing damping we have investigated this in detail in [22]. For nonvanishing damping the corresponding analysis is performed in [23].

Combining (69) and (77), the semiclassical equilibrium density matrix for inverse temperatures below $\theta_c$ may be written as

$$\rho_\theta(\bar{x},\bar{r}) = \frac{1}{Z}\frac{1}{\sqrt{-\Lambda}}\frac{1}{\sqrt{4\pi\theta^2}}\left(\prod_{n=1}^\infty \nu_n^2 u_n\right) \exp\left(\frac{i}{2}\Sigma_\theta(\bar{x},\bar{r})\right). \tag{78}$$

Here, $Z$ is a normalization constant and the minimal imaginary time action $\Sigma_\theta(\bar{x},\bar{r})$ is given by (69).

Having evaluated the effective action and the fluctuation integral for small $\epsilon$, we gain the time dependent semiclassical density matrix for small inverse temperatures up to inverse temperatures somewhat below $\theta_c$. We find

$$\rho(x_f, r_f, t) = \int dx_i\, dr_i\, d\bar{x}\, d\bar{r}\, J(x_f, r_f, t, x_i, r_i, \bar{x}, \bar{r})\, \lambda(x_i, r_i, \bar{x}, \bar{r}) \tag{79}$$

with the propagating function

$$J(x_f, r_f, t, x_i, r_i, \bar{x}, \bar{r}) = Z^{-1}\frac{1}{8\pi|A(t)|}\frac{1}{\sqrt{-\Lambda}}\frac{1}{\sqrt{4\pi\theta^2}}\left(\prod_{n=1}^\infty \nu_n^2 u_n\right)$$
$$\times \exp\left(\frac{i}{2}\Sigma_\theta(\bar{x},\bar{r}) + \frac{i}{2}\Sigma_t(x_f, r_f, t, x_i, r_i, \bar{x}, \bar{r})\right). \tag{80}$$

The actions $\Sigma_\theta(\bar{x},\bar{r})$ and $\Sigma_t(x_f, r_f, t, x_i, r_i, \bar{x}, \bar{r})$ are given by (69) and (74), respectively. Within the semiclassical approximation the above formula (79) gives the time evolution of the density matrix near the top of a potential barrier starting from an initial state with a deviation from thermal equilibrium described by the preparation function $\lambda(x_i, r_i, \bar{x}, \bar{r})$. We note that using the short time expansions of $A(t)$ and $S(t)$ (see [15]) one finds that (79) indeed reduces to (27) for $t \to 0$.

## V. STATIONARY FLUX SOLUTION

Now, we consider a system in a metastable state which may decay by crossing a potential barrier. We imagine that the system starts out from a potential well to the left of the barrier. Metastability means that the barrier height $V_b$ is much larger than other relevant energy scales of the system such as $k_B T$ and $\hbar\omega_0$, where $\hbar\omega_0$ is the excitation energy in the well of the inverted potential. The dynamics of the escape process may therefore be determined via a semiclassical calculation. In particular, we are interested here in the stationary flux over the barrier corresponding to a time independent escape rate. Naturally, a stationary flux does not exist for all times when the system is prepared at $t = 0$ in the metastable well. For short times preparation effects are important leading to a transient behavior. For very large times depletion of the state near the well bottom causes a flux decreasing in time. A plateau region with a quasi-stationary flux over the barrier exists only for intermediate times. We shall see that for high enough temperatures the flux state depends only on local properties of the barrier potential near the barrier top. The time evolution of an initial state in the metastable potential can then be calculated with the propagating function in (79). This will be done in this section. Firstly, in V A we introduce the inital preparation. Then, in V B we determine the stationary flux solution for high temperatures from the dynamics of the corresponding nonequilibrium state.

## A. Initial Preparation

The initial nonequilibrium state at time $t = 0$ is described by the preparation function

$$\lambda(x_i, r_i, \bar{x}, \bar{r}) = \delta(x_i - \bar{x})\delta(r_i - \bar{r})\Theta(-r_i) \tag{81}$$

so that the initial state is a thermal equilibrium state restricted to the left side of the barrier only. Then, according to (79), the dynamics is given by

$$\rho(x_f, r_f, t) = \int dx_i \, dr_i \, J(x_f, r_f, t, x_i, r_i, x_i, r_i) \, \Theta(-r_i). \tag{82}$$

To determine a stationary flux solution we seek for an approximately time independent state for intermediate times. As already mentioned, the real time dynamics near the barrier top is essentially given in terms of the functions $A(t)$ and $S(t)$ introduced in (71) and (72). Evaluating these functions for times larger than $1/\omega_R$ one gets to leading order an exponential growth according to

$$A(t) = -\frac{1}{2} \frac{1}{2\omega_R + \hat{\gamma}(\omega_R) + \omega_R \hat{\gamma}'(\omega_R)} \exp(\omega_R t) \tag{83}$$

and

$$S(t) = -\frac{1}{2} \cot(\frac{\omega_R \theta}{2}) \frac{1}{2\omega_R + \hat{\gamma}(\omega_R) + \omega_R \hat{\gamma}'(\omega_R)} \exp(\omega_R t). \tag{84}$$

These functions describe the unbounded motion at the parabolic barrier. Here, $\hat{\gamma}'(z)$ denotes the derivative of $\hat{\gamma}(z)$, and $\omega_R$ is the Grote-Hynes frequency [24] given by the positive solution of $\omega_R^2 + \omega_R \hat{\gamma}(\omega_R) = 1$. The corrections to (83) and (84) are exponentially decaying in time (see [15] for details). In the sequel we consider large times with $\omega_R t \gg 1$ so that the asymptotic behavior (83), (84) holds. We note that then terms in the effective action (74) like $\ddot{A}(t) - \dot{A}(t)^2/A(t)$ are exponentially small. For $\omega_R t \gg 1$ and $\bar{r} = r_i$, $\bar{x} = x_i$ (74) simplifies to read

$$\tilde{\Sigma}_t(x_f, r_f, t, x_i, r_i) = \Sigma_t(x_f, r_f, t, x_i, r_i, x_i, r_i) =$$
$$(x_f r_f + x_i r_i) \frac{\dot{A}(t)}{A(t)} + x_i r_f \frac{1}{2A(t)} - r_i x_i \left( \frac{\dot{A}(t)}{A(t)} + \frac{S(t)}{2\Lambda A(t)} \right)$$
$$+ \frac{i}{2} x_i^2 \left[ -\Omega + \frac{\Lambda}{4A(t)^2} \left( 1 - \frac{S(t)^2}{\Lambda^2} \right) \right]$$
$$+ i x_i x_f \frac{\Lambda \dot{A}(t)}{2A(t)^2} + \frac{i}{2} x_f^2 \left( \Omega + \Lambda \frac{\dot{A}(t)^2}{A(t)^2} \right) \tag{85}$$

where we have kept all terms with $x_i/A(t)$ since the relevant values of $x_i$ will be seen to become of order $A(t)$.

Hence, the density matrix (82) for times $\omega_R t \gg 1$ may be written as

$$\rho(x_f, r_f, t) = \int dx_i \, dr_i \, \tilde{J}(x_f, r_f, t, x_i, r_i) \, \Theta(-r_i) \tag{86}$$

where

$$\tilde{J}(x_f, r_f, t, x_i, r_i) = J(x_f, r_f, t, x_i, r_i, x_i, r_i) =$$
$$\frac{1}{Z} \frac{1}{8\pi|A(t)|} \frac{1}{\sqrt{-\Lambda}} \frac{1}{\sqrt{4\pi\theta^2}} \left(\prod_{n=1}^{\infty} \nu_n^2 u_n\right) \exp\left(\frac{i}{2}\tilde{\Sigma}(x_f, r_f, t, x_i, r_i)\right)$$
(87)

in which

$$\tilde{\Sigma}(x_f, r_f, t, x_i, r_i) = \Sigma_\theta(x_i, r_i) + \tilde{\Sigma}_t(x_f, r_f, t, x_i, r_i).$$
(88)

The two parts of this action are defined in (69) and (85).

### B. Stationary flux solution for high temperatures

We shall see that for high temperatures anharmonicities of the barrier potential are not relevant for the flux state. Since then the effective action is a bilinear function of the coordinates, the integrals in (86) are Gaussian and can be evaluated exactly.

The extremum of the effective action (88) with respect to $x_i$ and $r_i$ lies at

$$x_i^0 = -2\dot{A}(t)x_f + 2iA(t)\frac{r_f}{\Lambda}$$
$$r_i^0 = i\dot{S}(t)x_f + S(t)\frac{r_f}{\Lambda}.$$
(89)

Introducing the shifted coordinates

$$\hat{x}_i = x_i - x_i^0$$
$$\hat{r}_i = r_i - r_i^0,$$
(90)

the action may be written as

$$\tilde{\Sigma}(x_f, r_f, t, x_i, r_i) = \tilde{\Sigma}(x_f, r_f, t, x_i^0, r_i^0) + \hat{\Sigma}(\hat{x}_i, \hat{r}_i, t).$$
(91)

The first term is in fact independent of $t$ and coincides with the imaginary time action of the equilibrium density matrix, i.e.

$$\tilde{\Sigma}(x_f, r_f, t, x_i^0, r_i^0) = \Sigma_\theta(x_f, r_f)$$
(92)

where $\Sigma_\theta(x_f, r_f)$ was introduced in (69). For the time dependent second term in (91) one obtains

$$\hat{\Sigma}(\hat{x}_i, \hat{r}_i, t) = \frac{i}{2\Lambda}\left[\hat{r}_i^2 + i\frac{S(t)}{A(t)}\hat{r}_i\hat{x}_i - \frac{S(t)^2 - \Lambda^2}{4A(t)^2}\hat{x}_i^2\right].$$
(93)

With (92) and (93) the time dependent density matrix (82) may now be written as

$$\rho(x_f, r_f, t) = \rho_\theta(x_f, r_f)\, g(x_f, r_f, t)$$
(94)

where

$$\rho_\theta(x_f, r_f) = \frac{1}{Z} \frac{1}{\sqrt{-\Lambda}} \frac{1}{\sqrt{4\pi\theta^2}} \left( \prod_{n=1}^{\infty} \nu_n^2 \, u_n \right) \exp\left(\frac{i}{2} \Sigma_\theta(x_f, r_f)\right) \qquad (95)$$

is the equilibrium density matrix (78) for temperatures well above $T_c$, and

$$g(x_f, r_f, t) = \frac{1}{8\pi |A(t)|} \int d\hat{x}_i \, d\hat{r}_i \, \Theta(-\hat{r}_i - r_i^0) \exp\left(\frac{i}{2} \hat{\Sigma}(\hat{x}_i, \hat{r}_i, t)\right) \qquad (96)$$

is a "form factor" describing deviations from equilibrium.

To evaluate (96) we first perform the Gaussian integral over the $\hat{x}_i$-coordinate. Completing the square one finds

$$g(x_f, r_f, t) = \frac{1}{\sqrt{4\pi}} \sqrt{\frac{|\Lambda|}{S(t)^2 - \Lambda^2}} \int d\hat{r}_i \, \Theta(-\hat{r}_i - r_i^0) \exp\left(\frac{1}{4} \frac{\Lambda}{S(t)^2 - \Lambda^2} \hat{r}_i^2\right). \qquad (97)$$

With the transformation

$$\hat{r}_i = \frac{S(t)}{\Lambda} z - i\dot{S}(t) \, x_f \qquad (98)$$

the form factor reads

$$g(x_f, r_f, t) = \frac{1}{\sqrt{4\pi|\Lambda|}} \sqrt{\frac{S(t)^2}{S(t)^2 - \Lambda^2}}$$
$$\times \int dz \, \exp\left\{\frac{1}{4\Lambda} \frac{S(t)^2}{S(t)^2 - \Lambda^2} \left(z - i\Lambda \frac{\dot{S}(t)}{S(t)} x_f\right)^2\right\} \Theta\left(-\frac{S(t)}{\Lambda}[z + r_f]\right). \qquad (99)$$

¿From (84) we see that for $\omega_R \theta < \pi$ one has $S(t) < 0$. Now, since $\Lambda < 0$, the $\Theta$ function is nonvanishing only if

$$z < -r_f. \qquad (100)$$

Furthermore, for large times we have $\dot{S}(t)/S(t) = \omega_R$ and $S(t)^2/(S(t)^2 - \Lambda^2) = 1$ apart from exponentially small corrections. Thus, for $\omega_R t \gg 1$ the form factor becomes independent of time and reduces to

$$g_{\rm fl}(x_f, r_f) = \frac{1}{\sqrt{4\pi|\Lambda|}} \int_{-\infty}^{-r_f} dz \, \exp\left\{\frac{1}{4\Lambda} (z - i\Lambda\omega_R x_f)^2\right\}. \qquad (101)$$

This combines with (94) to yield the stationary flux solution

$$\rho_{\rm fl}(x_f, r_f) = \rho_\theta(x_f, r_f) \frac{1}{\sqrt{4\pi|\Lambda|}} \int_{-\infty}^{u} dz \, \exp\left(-\frac{z^2}{4|\Lambda|}\right) \qquad (102)$$

where

$$u = -r_f + i|\Lambda|\omega_R x_f. \tag{103}$$

To illustrate the result (102) the diagonal part $g_{fl}(0,q)$ of the form factor (101) is shown in Fig. 1 as a function of the scaled coordinate $q$ for ohmic damping with $\hat{\gamma} = 3$ and three different values of the inverse temperature $\theta$. One sees that with increasing $\theta$ the width of the nonequilibrium state becomes smaller due to the fact that energy fluctuations of the Brownian particle decrease for lower temperatures. We note that for $\hat{\gamma} = 3$ the function $\Lambda$ vanishes at the critical inverse temperature $\theta_c = 5.79\ldots$.

Finally, let us discuss the range of validity of the time independent flux solution (102). Firstly, in view of (93), the integral over the $\hat{x}_i$-coordinate in the expression (96) for the form factor is dominated by values of $\hat{x}_i$ of order $\sqrt{-\Lambda}$ or smaller. After performing the $\hat{x}_i$-integration, the relevant coordinate $\hat{r}_i/S(t)$ in the remaining integral (97) becomes of order $1/\sqrt{-\Lambda}$ or smaller. For high enough temperatures $\hat{x}_i$ and $\hat{r}_i/S(t)$ are then of order 1 or smaller which is consistent with the harmonic approximation. However, with increasing inverse temperature $|\Lambda|$ decreases and the relevant values of $\hat{r}_i/S(t)$ become larger. Therefore, the harmonic approximation of the barrier potential which leads to the Gaussian integrals in (96) is valid only for high enough temperatures. We will show in [23] that anharmonicities of the potential barrier become essential for temperatures near the critical temperature $T_c$ where the first caustic in the harmonic potential arises.

Secondly, as mentioned above, the flux state (102) was found only for $\omega_R t \gg 1$. There is also an upper bound of time due to the fact that the flux solution was obtained by the harmonic approximation of the potential. To estimate this upper bound we consider coordinates within the width of the diagonal part of the flux solution (102) which is of order $\sqrt{|\Lambda|}$. Anharmonic terms of the potential (20) are negligible when $\epsilon^2 q^2 \ll 1$, i.e. $q_d \ll q_a$ where $q_d$ is the dimensional coordinate. Hence, inserting the $r_f$-dependent term of the shift $r_i^0$ in (89) for values of $r_f$ of order $\sqrt{|\Lambda|}$ one gains the condition $\epsilon^2 S(t)^2/|\Lambda| \ll 1$. According to (84) for large times $S(t)$ becomes of order $(\omega_R/2)\cot(\omega_R\theta/2)\exp(\omega_R t)$. Furthermore, as far as orders of magnitudes are concerned $(\omega_R/2)\cot(\omega_R\theta/2)$ is of same order as $|\Lambda|$ for all temperatures well above $T_c$. This can be seen by estimating $\Lambda$ from (55) in various limits. As a consequence, anharmonic terms in the potential can be neglected only if $\exp(\omega_R t) \ll 1/\epsilon\sqrt{|\Lambda|}$. Combining the two bounds we see that a stationary flux solution exists within the plateau region

$$1 \ll \exp(\omega_R t) \ll \frac{1}{\epsilon\sqrt{|\Lambda|}} \approx \frac{1}{\epsilon}\sqrt{\frac{2\tan(\omega_R\theta/2)}{\omega_R}}. \tag{104}$$

For very high temperatures $\theta \ll 1$, where $|\Lambda| \approx 1/\theta$, the upper bound of time reduces to $\exp(\omega_R t) \ll \sqrt{\theta}/\epsilon$ which can be written as $\exp(\omega_R t) \ll q_a\sqrt{\beta m\omega_0^2/2}$ in dimensional units. For lower temperatures, where $|\Lambda|$ is of order 1, one obtains $\exp(\omega_R t) \ll q_a/q_0$ in dimensional units. We note that (104) holds only for inverse temperatures $\theta < \theta_c$ where $|\Lambda|$ is finite. In summary the flux solution (102) is valid for high enough temperatures and times within the range (104).

The result (104) has a simple physical interpretation. The exact propagating function of the nonlinear potential gives the probability amplitude to reach arbitrary coordinates $(x_f, r_f)$ when starting from coordinates $(x_i, r_i)$. Now, consider endpoints $(x_f, r_f)$ near the barrier top. Clearly, the main contribution to the corresponding probability amplitude then comes from paths with energy of order of the potential energy at the barrier top. Other paths either have smaller energy and do not reach the barrier top or much larger energy and are therefore exponentially suppressed. However, a trajectory starting at $t=0$ far away from the barrier top in the anharmonic range of the potential with an energy of order of the potential energy at the barrier top needs at least a time of order $\ln(q_a/\Delta q_d)$ to reach a region of order $\Delta q_d$ about the barrier top. Now, in dimensional units the flux state has a width of order $\Delta q_d = q_0\sqrt{|\Lambda|}$. Hence, for times sufficiently smaller than $\ln(q_a/\Delta q_d) = \ln(1/\epsilon\sqrt{|\Lambda|})$ the probability to reach endpoints near the barrier top is basically determined by trajectories starting also near the barrier top. The upper bound in (104) gives the time where trajectories starting far from the barrier region contribute essentially to the nonstationary state.

## VI. MATCHING TO EQUILIBRIUM STATE IN THE WELL AND DECAY RATE

With (102) we have found an analytic expression for a stationary flux state of a metastable system for temperatures above $T_c$. The function $\rho_{\text{fl}}(x_f, r_f)$ describes the reduced density matrix of the system for coordinates in the barrier region and for times within the plateau region (104). This nonequilibrium state depends on local properties of the metastable potential near the barrier top only. On the other hand the metastable system is in thermal equilibrium near the well bottom. This means that the flux solution must reduce to the thermal equilibrium state for coordinates $q_f$, $q'_f$ on the left side of the barrier sufficiently far from the barrier top but at distances much smaller than $\epsilon^{-1}$ which is the typical distance from the barrier top to the well bottom. In this section we verify this condition for the high temperature nonequilibrium state (102) and use the result to derive the decay rate of the metastable state in the well.

### A. Matching of flux solution to equilibrium state

For coordinates $q_f$, $q'_f$ to the left of the barrier the form factor (102) has to approach 1 as one moves away from the barrier top. Let us investigate the region of coordinates $x_f$, $r_f$ where

$$\left| 1 - \frac{1}{\sqrt{4\pi|\Lambda|}} \int_{-\infty}^{u} dz\, \exp\left(-\frac{z^2}{4|\Lambda|}\right) \right| \lesssim \exp(-1/\epsilon^\mu) \ll 1. \tag{105}$$

Here, $u = u(x_f, r_f)$ is given by (103) and the exponent $\mu$ is positive. In the semiclassical limit $\mu$ can be small. Now, in the halfplane $r_f < 0$ of the $x_f r_f$-plane the condition (105) defines a region which may be delineated by the relation

$$|x_f| \lesssim \left( \frac{r_f^2}{\omega_R^2 \Lambda^2} - \frac{4}{\omega_R^2 |\Lambda| \epsilon^\mu} \right)^{1/2}. \tag{106}$$

On the other hand, the equilibrium density matrix (95) is nonvanishing essentially only for

$$|x_f| \lesssim \frac{|r_f|}{\sqrt{|\Lambda|\Omega}}. \tag{107}$$

Now, in order that the flux solution matches to the equilibrium state within the region of coordinates where (95) and (96) are both valid, (106) and (107) must overlap in the $x_f r_f$-plane for sufficiently large $r_f < 0$ but $|r_f| \ll \epsilon^{-1}$. Thereby, one has to take into account that due to the definitions of $r_f$ and $x_f$ in (22) one has $|x_f| \leq 2|r_f|$ for $q_f, q'_f < 0$. For high temperatures this latter relation is always fulfilled by virtue of (107). Hence, from the relations (106) and (107) one obtains

$$|\Lambda|\epsilon^{2-\mu} \ll 1 - \omega_R^2 \frac{|\Lambda|}{\Omega} \tag{108}$$

where the $\mu$-dependence may be disregarded, since we may choose $\mu \ll 1$.

The flux solution (102) describes the density matrix of the metastable system in the barrier region only when (108) is satisfied. This condition depends on temperature, anharmonicity parameter, and damping. We note that in dimensionless units $\epsilon^{-2}$ is the typical barrier height with respect to the well bottom.

So far, we have written all results in terms of dimensionless quantities according to the definitions in subsection II C, except for a few formulas where the change to dimensional units was mentioned explicitly in the text. In order to facilitate a comparison with earlier results we shall *return to dimensional units* for the remainder of this section. Then, we obtain from (108) in dimensional units the condition

$$|\Lambda| \ll \frac{V_b}{\hbar \omega_0^2} \left(1 - \omega_R^2 \frac{|\Lambda|}{\Omega}\right). \tag{109}$$

Here, according to (55) and (70) one has in dimensional units

$$\Lambda = \frac{1}{\hbar \beta} \sum_{n=-\infty}^{\infty} \frac{1}{\nu_n^2 + |\nu_n|\hat{\gamma}(|\nu_n|) - \omega_0^2} \tag{110}$$

and

$$\Omega = \frac{1}{\hbar \beta} \sum_{n=-\infty}^{\infty} \frac{|\nu_n|\hat{\gamma}(|\nu_n|) - \omega_0^2}{\nu_n^2 + |\nu_n|\hat{\gamma}(|\nu_n|) - \omega_0^2}, \tag{111}$$

where $\omega_0$ denotes the oscillation frequency at the barrier top, $V_b$ the barrier height with respect to the well bottom, and the

$$\nu_n = \frac{2\pi n}{\hbar \beta} \tag{112}$$

are the Matsubara frequencies. The dimensional Grote–Hynes frequency is determined by the largest positive root of the equation

$$\omega_R^2 + \omega_R \hat{\gamma}(\omega_R) - \omega_0^2 = 0. \tag{113}$$

¿From a physical point of view (109) defines the region where the influence of the heat bath on the escape dynamics is strong enough to destroy coherence on a length scale smaller than the scale where anharmonicities becomes important. Only then are nonequilibrium effects localized in coordinate space to the barrier region. For very high temperatures, i.e. in the classical region $\omega_0\hbar\beta \ll 1$, one has $|\Lambda| \approx 1/\omega_0^2\hbar\beta$ and $\Omega \approx 1/\hbar\beta$. Then, one obtains from (109) for Ohmic dissipation with $\hat{\gamma}(\omega) = \gamma$ the well-known Kramers condition [2] $k_B T \omega_0 / V_b \ll \gamma$ where we have used $1 - \omega_R^2/\omega_0^2 \approx \gamma/\omega_0$ for small damping. When the temperature is lowered $|\Lambda|$ decreases and the range of damping where the flux solution (102) is valid becomes larger.

To make the condition (109) more explicit, especially for lower temperatures, one has to specify the damping mechanism. As an example we consider a Drude model with $\gamma(t) = \gamma\omega_D \exp(-\omega_D t)$. Clearly, in the limit $\omega_D \gg \omega_0, \gamma$ the Drude model behaves like an Ohmic model except for very short times of order $1/\omega_D$. The Laplace transform of $\gamma(t)$ then reads

$$\hat{\gamma}(z) = \frac{\gamma\omega_D}{\omega_D + z}. \tag{114}$$

Since the condition (109) is relevant only for small damping strength, it suffices to evaluate (109) in leading order in $\gamma$. Then, from (113) one gets

$$\omega_R = \omega_0 - \frac{\gamma}{2}\frac{\omega_D}{\omega_D + \omega_0} + \mathcal{O}(\gamma^2). \tag{115}$$

Expanding $\Lambda$ given in (110) up to first order in $\gamma$ yields

$$\Lambda(\gamma) = -\frac{1}{2\omega_0}\cot(\omega_0\hbar\beta/2) + \gamma\Lambda'(0) + \mathcal{O}(\gamma^2) \tag{116}$$

where

$$\Lambda'(0) \equiv \left.\frac{\partial\Lambda}{\partial\gamma}\right|_{\gamma=0} = -\frac{1}{\hbar\beta}\sum_{n=-\infty}^{\infty}\frac{|\nu_n|\omega_D}{(\nu_n^2 - \omega_0^2)^2(\omega_D + |\nu_n|)}. \tag{117}$$

Correspondingly, for $\Omega$ we gain from (111)

$$\Omega(\gamma) = \frac{\omega_0}{2}\cot(\omega_0\hbar\beta/2) + \gamma\Omega'(0) + \mathcal{O}(\gamma^2) \tag{118}$$

where

$$\Omega'(0) \equiv \left.\frac{\partial\Omega}{\partial\gamma}\right|_{\gamma=0} = \frac{1}{\hbar\beta}\sum_{n=-\infty}^{\infty}\frac{|\nu_n|^3\omega_D}{(\nu_n^2 - \omega_0^2)^2(\omega_D + |\nu_n|)}. \tag{119}$$

Combining (115)–(119) the condition (109) takes the form

$$\gamma \gg \frac{\omega_D + \omega_0}{\omega_D}\frac{\hbar\omega_0^2}{2V_b\tan(\omega_0\hbar\beta/2)}\frac{1}{1 + 2\kappa\tan(\omega_0\hbar\beta/2)(\omega_D + \omega_0)/\omega_D} \tag{120}$$

where

$$\kappa = \Omega'(0) + \omega_0^2 \Lambda'(0) = \frac{1}{\hbar\beta} \sum_{n=-\infty}^{\infty} \frac{|\nu_n|\omega_D}{(\nu_n^2 - \omega_0^2)(\omega_D + |\nu_n|)}. \tag{121}$$

Let us discuss the relation (120) in the limit $\omega_D \gg \omega_0, \gamma$. For very high temperatures, i.e. in the classical region $\omega_D \hbar\beta \ll 1$, the last factor in (120) approaches 1 and (120) reduces to the Kramers condition $\gamma \gg \omega_0/V_b\beta$. For lower temperatures where $\omega_D \hbar\beta \gg 1$, but $\omega_0 \hbar\beta < \pi$, the coefficient $\kappa$ becomes large. In leading order we have [25]

$$\kappa = \frac{\ln(\omega_D \hbar\beta)}{\pi}. \tag{122}$$

Hence, (120) gives

$$\gamma \gg \frac{\hbar\omega_0^2}{2V_b \tan(\omega_0 \hbar\beta/2)} \frac{1}{1 + 2\ln(\omega_D \hbar\beta)\tan(\omega_0 \hbar\beta/2)/\pi}. \tag{123}$$

Clearly, the range of $\gamma$ where the result (102) is valid increases for lower temperatures as already stated above. Finally, we note that the conditions (109) and (120) hold only for temperatures above $T_c$ where $|\Lambda|$ is finite.

## B. Decay rate of metastable state

Clearly, the flux solution (102) contains all relevant informations about the nonequilibrium state. Specifically, the steady-state decay rate $\Gamma$ of the metastable state is given by the momentum expectation value in the flux state at the barrier top. In dimensional units one has

$$\Gamma = \frac{1}{Z}\frac{1}{2M}\langle \hat{p}\delta(\hat{q}) + \delta(\hat{q})\hat{p}\rangle_{\rm fl} \tag{124}$$

where the expectation value $\langle \cdot \rangle_{\rm fl}$ is calculated with respect to the dimensional stationary flux solution. Here, the normalization constant $Z$ is determined by the matching of the flux solution onto the thermal equilibrium state inside the well. From (124) one has in coordinate representation

$$\Gamma = \frac{J_{\rm fl}}{Z} = \left( \frac{\hbar}{iM}\frac{\partial}{\partial x_f} \rho_{\rm fl}(x_f, 0) \right)\bigg|_{x_f=0}. \tag{125}$$

Here, $J_{\rm fl}$ is the total probability flux at the barrier top $q = 0$. Since the essential contribution to the population in the well comes from the region near the well bottom, $Z$ can be approximated by the partition function of a damped harmonic oscillator with frequency $\omega_w$ at the well bottom, i.e.

$$Z = \frac{1}{\omega_w \hbar\beta} \left( \prod_{n=1}^{\infty} \frac{\nu_n^2}{\nu_n^2 + |\nu_n|\hat{\gamma}(|\nu_n|) + \omega_w^2} \right) \exp(\beta V_b). \tag{126}$$

Here, $V_b$ denotes the barrier height with respect to the well bottom in dimensional units. Note that the potential was set to 0 at the barrier top. Inserting (102) for $r_f = 0$ and (126) into (125) gives in dimensional units

$$\Gamma = \frac{\omega_w}{2\pi} \frac{\omega_{\rm R}}{\omega_0} \left( \prod_{n=1}^{\infty} \frac{\nu_n^2 + |\nu_n|\hat{\gamma}(|\nu_n|) + \omega_w^2}{\nu_n^2 + |\nu_n|\hat{\gamma}(|\nu_n|) - \omega_0^2} \right) \exp(-\beta V_b). \tag{127}$$

We recall that the dimensional Grote-Hynes frequency $\omega_{\rm R}$ is given by the positive solution of $\omega_{\rm R}^2 + \omega_{\rm R}\hat{\gamma}(\omega_{\rm R}) = \omega_0^2$. As is apparent from the Arrhenius exponential factor, the rate (127) describes thermally activated transitions across the barrier where the prefactor takes into account quantum corrections. The above decay rate is well-known from imaginary time path integral methods [12] and equivalent approaches [1,11,26]. These methods consider static properties of the metastable state only. Here, the expression (127) was derived as a result of a dynamical calculation. This avoids the crucial assumption of the imaginary time method that the imaginary part of the analytically continued free energy of the unstable system may be interpreted as a decay rate. In this sense, we have put the result (127) on firmer grounds. Moreover, we have obtain the condition (108) on the damping range where (127) is valid.

## VII. CONCLUSIONS

Starting from a path integral representation of the density matrix of a dissipative quantum system with a potential barrier we have derived an evolution equation for the density matrix in the vicinity of the barrier top. This equation was shown to have a quasi-stationary nonequilibrium solution with a constant flux across the potential barrier. In this paper we have restricted ourselves to the temperature region where the barrier is crossed primarily by thermally activated processes. In this region the flux solution was shown to be independent of anharmonicities of the barrier potential. This nonequilibrium state generalizes the well–known Kramers solution of the classical Fokker–Planck equation to the region where quantum corrections are relevant.

On one side of the barrier the flux solution approaches an equilibrium state as one moves away from the barrier top. We have obtained a condition on the damping strength which ensures that the equilibrium state is reached within the range of validity of the harmonic approximation for the barrier potential. Again this condition can be shown to be a generalization of a corresponding condition known from classical rate theory. The flux solution can then be matched to the equilibrium state in the metastable well which yields the proper normalization of the flux across the barrier. In particular, we have used the normalized flux state to determine the decay rate of the metastable state. The result was shown to be identical with the well–known rate formula for thermally activated decay in the presence of quantum corrections as derived by purely thermodynamic methods. The dynamical theory presented here gives in addition a criterion for the range of damping parameters where this result is valid.

A particularly interesting feature of the general approach advanced in this article is the fact that it can be extended both to lower temperatures and smaller damping. This will be the subject of subsequent work.

## ACKNOWLEDGMENTS

Financial support was provided by the Deutsche Forschungsgemeinschaft (Bonn) through SFB237. One of us (G.-L.I.) was also supported additionally through a Heisenberg fellowship.

# REFERENCES

[1] P. Hänggi, P. Talkner, and M. Borkovec, Rev. Mod. Phys. **62**, 251 (1990).
[2] H. A. Kramers, Physica **7**, 284 (1948).
[3] A. O. Caldeira and A. J. Leggett, Phys. Rev. Lett. **46**, 211 (1981); A. O. Caldeira and A. J. Leggett, Ann. Phys. (USA) **149**, 374 (1983); **153**, 445(E) (1984).
[4] A. I. Larkin and Yu. N. Ovchinnikov, Pis'ma Zh. Eksp. Teor. Fiz. **37**, 322 (1983) [ Sov. Phys.-JETP **37** , 382 (1983)]; A. I. Larkin and Yu. N. Ovchinnikov, Zh. Eksp. Teor. Fiz. **86**, 719 (1984) [ Sov. Phys.-JETP **59**, 420 (1984)].
[5] H. Grabert, U. Weiss, and P. Hänggi, Phys. Rev. Lett. **52**, 2193 (1984); H. Grabert and U. Weiss, Phys. Rev. Lett. **53**, 1787 (1984).
[6] J. S. Langer, Ann. Phys. (N.Y.) **41**, 108 (1967).
[7] M. Stone, Phys. Lett. **67B**, 186 (1977).
[8] C. G. Callan and S. Coleman, Phys. Rev. D **16**, 1762 (1977). S. Coleman, in: The Whys of Subnuclear Physics, edited by A. Zichichi (Plenum, New York, 1979).
[9] W. H. Miller, J. Chem. Phys. **62**, 1899 (1975); W. H. Miller, Adv. Chem. Phys. **25**, 69 (1974).
[10] E. Pollak, Phys. Rev. A **33**, 4744 (1986); E. Pollak, Chem. Phys. Lett. **177**, 178 (1986).
[11] P. Hänggi and W. Hontscha, J. Chem. Phys. **88**, 4094 (1988); P. Hänggi and W. Hontscha, Ber. Bunsenges. Phys. Chem. **95**, 379 (1991).
[12] H. Grabert, P. Olschowski, and U. Weiss, Phys. Rev. **B 36**, 1931 (1987).
[13] M. H. Devoret, D. Esteve, C. Urbina, J. Martinis, A. Cleland, and J. Clarke, in: Quantum Tunneling in Solids, edited by Yu. Kagan and A. J. Leggett (Elsevier, New York, 1992).
[14] R. P. Feynman and F. L. Vernon, Ann. Phys. (N.Y.) **243**, 118 (1963).
[15] H. Grabert, P. Schramm, and G.-L. Ingold, Phys. Rep. **168**, 115 (1988).
[16] U. Weiss, Quantum Dissipative Systems (World Scientific, Singapore, 1993).
[17] G.-L. Ingold, Ph.D. thesis (Stuttgart, 1988).
[18] H. Hofmann and G.-L. Ingold, Phys. Lett. B **264**, 253 (1991).
[19] P. Ullersma, Physica **32**, 27, 56, 74, 90 (1966).
[20] R. P. Feynman and A. P. Hibbs, Quantum Mechanics and Path Integrals (McGraw-Hill, New York, 1965); R. P. Feynman, Statistical Mechanics (Benjamin, New York, 1972).
[21] L. S. Schulman, Techniques and Applications of Path Integrals (Wiley, New York, 1981).
[22] J. Ankerhold and H. Grabert, Physica **A188**, 568 (1992).
[23] J. Ankerhold and H. Grabert, Dissipative Quantum Systems with Potential Barrier. II. Dynamics Near the Barrier Top, to be published.
[24] R. F. Grote and J. T. Hynes, J. Chem. Phys. **73**, 2715 (1980); P. Hänggi and F. Mojtabai, Phys. Rev. A **26**, 1168 (1982).
[25] H. Grabert, U. Weiss, and P. Talkner, Z. Phys. B **55**, 87 (1984).
[26] P. G. Wolynes, Phys. Rev. Lett. **47**, 968 (1981).

FIGURES

FIG. 1. Diagonal part $g_{fl}(0, q)$ of the form factor of the stationary flux solution (102) as a function of the scaled coordinate $q$ for ohmic damping with $\hat{\gamma} = 3$ and different values of the inverse temperature $\theta$. The dashed line corresponds to $\theta = 0.1$, the dotted–dashed line to $\theta = 0.5$, and the solid line to $\theta = 3.0$.

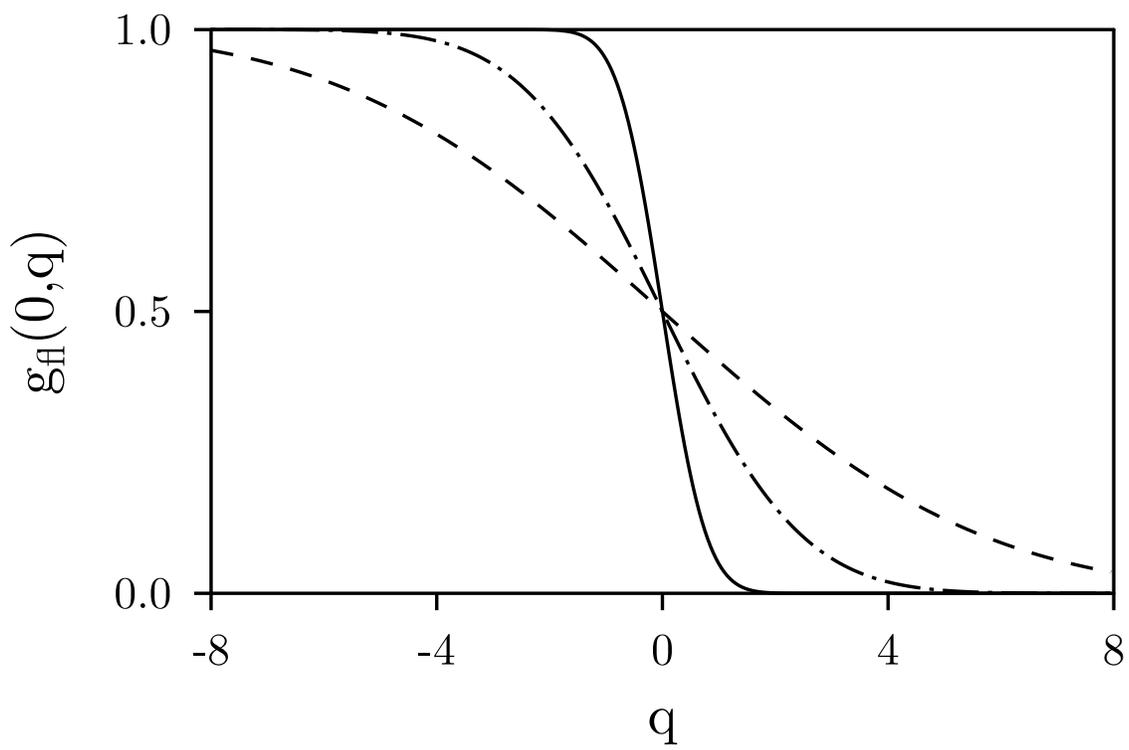

Figure 1

Ankerhold et al., Dissipative Quantum Systems ...